# Conformational composition of propanol in gaseous state and in matrix isolation

Iryna Doroshenko (1), Laziz Meyliev (2), Bahrom Kuyliev (2)

((1) Taras Shevchenko National University of Kyiv, Kyiv, Ukraine, (2) Karshi State University, Karshi, Uzbekistan)

Conformational analysis of the experimentally recorded IR absorption spectra of propanol in gaseous state and in a low-temperature argon matrix at 20 K and at 35 K was carried out for different spectral ranges. It showed that the conformational composition of the samples in gas and in matrix isolation is different. In gaseous propanol Gt conformers predominate, while in matrix isolation the most energetically favorable form is Tg conformer, which prevails in percentage at both considered temperatures of an Ar matrix.

**Introduction**

Propanol is one of the most useable alcohols for different applications, so it is widely studied both by experimental and theoretical methods. It is known that molecules of monohydric alcohols can form different cluster structures due to hydrogen bonding [1-3]. Moreover, each molecule of alcohol with more than one atom of carbon can be found in different conformations, which are formed as a result of rotation of atoms around chemical bonds [4-6]. In the case of propanol (structural formula $CH_2$-$CH_3$-$CH_3$-OH), the molecule of which has two structural dihedrals, nine stable configurations exist: one plane structure and four pairs of enantiomers (or mirror-image pairs). Since enantiomers have similar energy and optical properties, the difference between them is usually neglected. As a rule, propanol conformers are designated by a symbolic title using a generally accepted scheme of dihedral angles CCCO+CCOH via big+small letters for trans- (T, t), gauche- (G, g) and gauche'- (G', g') conformers [6, 7].

Conformational properties of propanol were studied in supersonic jets [6, 8] and cryogenic matrices [6, 9], in gas and liquid phases [7], in $CCl_4$ solutions at low concentrations [10]. The aim of the presented work is to analyze experimentally registered IR absorption spectra of propanol in gaseous state and in matrix isolation, and to compare conformational composition of the samples studied.

**Experimental**

Liquid propanol with purity > 99.9 from Fluka was used for experimental investigations. Atmospheric oxygen and nitrogen were removed from the alcohol by five freeze-pump-thaw cycles. Gas phase samples were obtained by the natural evaporation in vacuum process from the liquid surface.

FTIR spectra of propanol in gaseous state were measured using conventional single pass 10 cm gas cell equipped with KBr windows with the resolution 0.5 cm$^{-1}$ using deuterated triglycine sulfate (DTGS) detector. The gas cell was filled with an alcohol using standard vacuum technique. Pressure of the alcohol in the cell was set close to its saturated pressure at 20º C: 2 mbar.

**Conformational analysis of gaseous propanol**

*Spectral region of OH stretching vibrations*

Fig. 1 presents the spectral band of the stretching vibrations of free hydroxyl groups (3600 – 3750 cm$^{-1}$) of gaseous propanol. A complicated shape of this band is due to the vibrations of different conformers of propanol molecule as well as vibrations of non-bonded hydroxyl groups of dimers and bigger clusters [11-13].

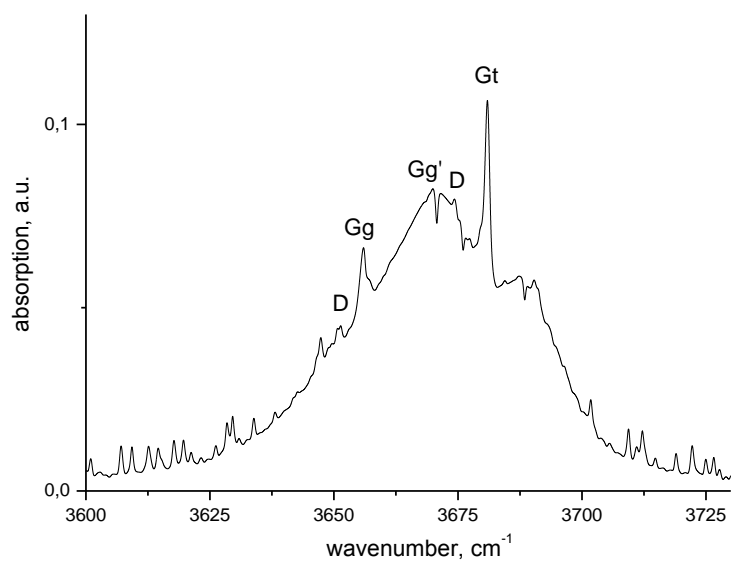

Fig.1 – IR absorption band of gaseous propanol corresponding to the stretching vibrations of free OH groups

In [14] the following assignment of the absorption bands to different propanol conformers was made: 3682 cm$^{-1}$ – Gt conformer, 3679 cm$^{-1}$ – Tt conformer, 3669 cm$^{-1}$ – Gg', 3660 cm$^{-1}$ – Tg and 3657 cm$^{-1}$ – Gg. As is seen from Fig.1, one can distinguish an intense band at 3682 cm$^{-1}$, which can be assigned to vibrations of Gt conformer. In addition, there are bands at 3657 and 3669 cm$^{-1}$, which can be assigned to conformers Gg and Gg'. IR absorption bands of Tt and Tg conformers are not observed in this part of the spectrum, or their intensity is too low. Therefore, we can draw the conclusion that conformations with gauche- orientation of C – C bond – Gt, Gg and Gg' – dominate in the studied sample of gaseous propanol.

Moreover, spectral bands of molecules, which are proton acceptors in dimers, with frequencies 3654, 3671 and 3675 cm$^{-1}$ were found in [14]. All these three bands were found in our sample, too. In fig. 1, D denotes the dimer bands. Besides these identified bands there are other ones (for example, 3647 cm$^{-1}$), which can belong to vibrations of free OH groups in bigger clusters – trimers and so on.

*Spectral region 800 – 1300 cm$^{-1}$*

In the spectral region from 800 to 1300 cm$^{-1}$ the stretching vibrations of a carbon skeleton of the propanol molecule (i.e. C – C and C – O bonds) are located. Fig. 2 presents FTIR spectrum of gaseous propanol in the corresponding range.

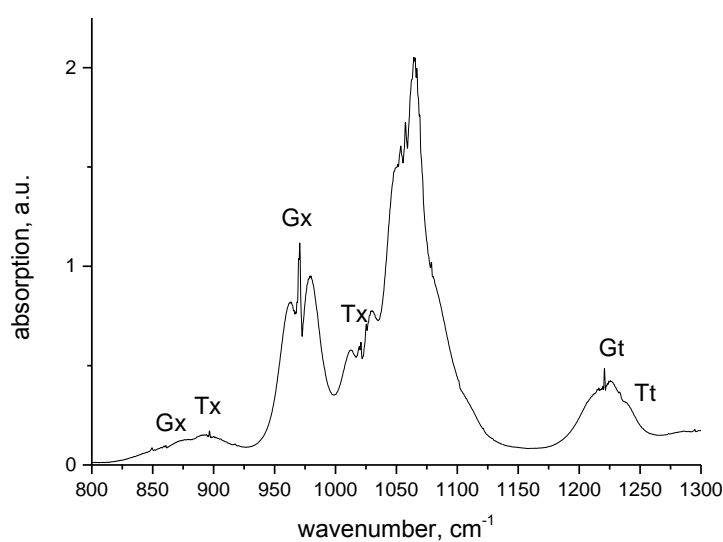

Fig.2 – FTIR spectrum of gaseous propanol in the range 800 – 1300 cm$^{-1}$

The stretching C – C vibrations of conformers with trans-configuration in respect to C – C bond (they are denoted as Tx) have frequencies 881 and 885 cm$^{-1}$ [14], and the corresponding vibrations of conformers of Gx group are located in the range 850 – 860 cm$^{-1}$. In the experimentally registered spectrum of gaseous propanol there are vibrations at these frequencies, but they are of low intensity.

In the spectrum presented in Fig. 2 the most intense absorption is observed at 1065 cm$^{-1}$. This band corresponds to a more high-frequency part of a doublet appearing at interaction of the stretching C – C and C – O vibrations [14]. However, this band gives few information about dominating conformations, since calculated frequencies for all five conformers are located in the range 1050 – 1100 cm$^{-1}$. The second, low-frequency component of the doublet, is more informative: vibrations of Tx conformers are observed at 1014 and 1025 cm$^{-1}$, while vibrations of Gx conformers are located between 965 and 972 cm$^{-1}$ [14]. Therefore, the observed absorption maximum at 1020 cm$^{-1}$ can be assigned to Tx vibrations, and the band at 972 cm$^{-1}$ – to Gx vibrations. The comparison of intensities of these two bands allows concluding that Gx conformers dominate in the investigated sample, which agree with the conclusions drawn from the analysis of the spectral region of the stretching OH vibrations.

In addition to skeleton vibrations, in this region there are also bending OH vibrations. According to [14], bending OH vibrations in Gt conformer manifest at 1222 cm$^{-1}$, and the corresponding vibrations of Tt conformer – at 1235 cm$^{-1}$. In the IR spectrum in Fig. 2 one can notice both of these bands, and an intensity of Gt band is higher than that of Tt band, indicating domination of Gt form over Tt one.

*Spectral region of stretching C – H vibrations*

Fig. 3 presents IR absorption spectrum of gaseous propanol in the spectral range of the stretching C – H vibrations. According to [15], the absorption band near 2890 cm$^{-1}$ corresponds to the symmetric stretching C-H vibrations in CH$_3$ group, and the band at 2972 cm$^{-1}$ – to antisymmetric stretching C-H vibrations in CH$_3$ group. Between them, the symmetric stretching C-H vibrations of C$_\beta$H$_2$ group are located. According to the assignment from [16 - 18], the band at 2942 cm$^{-1}$ belongs to the conformer Tg, the band at 2952 cm$^{-1}$ – to the conformer Gg. It means that the most intense vibrations in this region correspond to conformers, which have OH group in gauche- position. Absorption frequencies of conformers with hydroxyl group in trans- position are lower by several tens of wavenumbers. However, in the registered spectrum absorption at these frequencies is low.

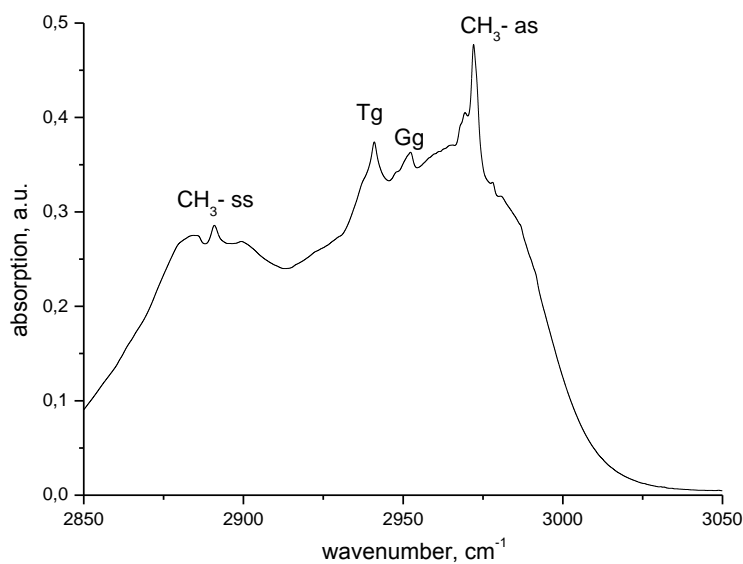

Fig. 3 – IR spectrum of gaseous propanol in the region of the stretching C – H vibrations

**Conformational analysis of propanol in matrix isolation**

Matrix isolation in low-temperature matrices of inert gases allows registering vibrational spectra of individual molecules or small clusters without their interaction with environment. Moreover, in contrast to the spectra of a substance in the gas phase, the vibrational bands in the obtained spectra are not complicated by the rotational structure. This fact allows making a detailed assignment of registered bands to different conformers or clusters of different sizes. In this section, IR spectra of propanol in low-temperature argon matrix are considered. For samples preparation, propanol vapor was mixed with argon in ratio 1:1000 (one propanol molecule per 1000 argon atoms), then the obtained gas mixture was deposited at a cooled to 20K window. As a result, one can register IR spectra of propanol molecules or small clusters and analyze conformational composition of propanol in matrix isolation.

*Spectral region of stretching OH vibrations*

Fig. 4 presents IR spectrum of propanol trapped in an Ar matrix at 20 K in the spectral region 3640 – 3680 cm$^{-1}$. In this part of the spectrum, the stretching vibrations of non-bonded hydroxyl group are manifested. These non-bonded OH groups can belong to monomers and proton acceptors in dimers as well as bigger chain clusters.

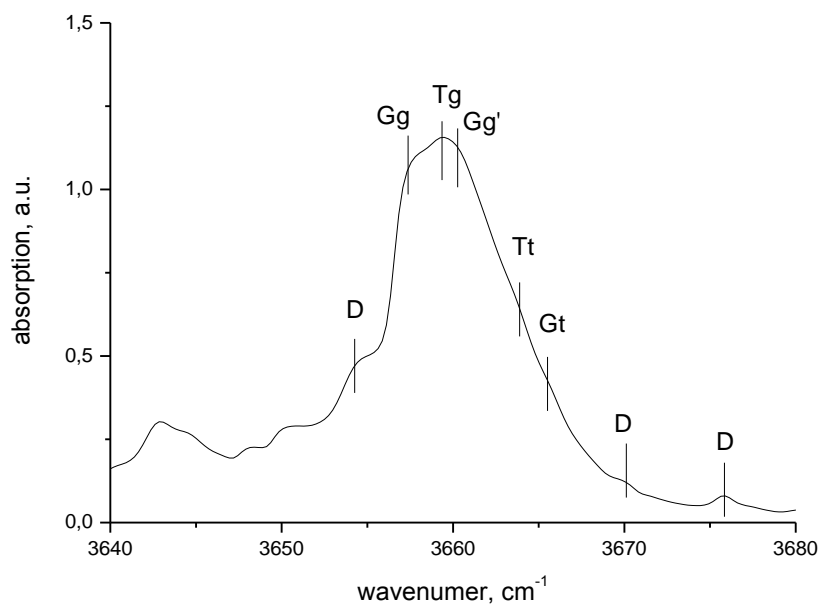

Fig. 4 – IR spectrum of propanol trapped in an Ar matrix at 20 K in the spectral region of stretching OH vibrations

As is seen, the spectral band in fig. 4 has a complex contour with a number of maxima. The maxima at 3657, 3659, 3660, 3663, and 3665 cm$^{-1}$ are assigned, respectively, to the vibrations of Gg, Tg, Gg', Tt and Gt conformers. This assignment was made according to the results of IR absorption of propanol conformers in an Ar matrix in this spectral region [19]. Analyzing relative intensities of the spectral bands corresponding to the stretching OH vibrations of five different propanol conformers one can say that conformers with gauche-configuration of OH group (i.e. Gg, Tg and Gg') dominate in the investigated sample. The intensity of the bands corresponding to Tt and Gt conformers is essentially lower, indicating that the number of these conformers (with OH group in trans- position) in the sample is much less. It is interesting to note, that in the gas phase the domination of Gt conformer was observed in this spectral region. Apparently, during the cooling of the gas mixture and the deposition of the matrix, a transition from one form to another occurred, and the conformational composition of the sample changed. This is confirmed by the results of our quantum-chemical calculations [20], according to which the lowest energy barrier (about 3.2 kJ/mol) is observed between Gt and Gg 'conformers, that is, the transition from Gt conformation to Gg' conformation is the most probable.

In addition to the vibrations of propanol monomers, spectral bands of OH stretching vibrations of free (i.e., not involved in the formation of hydrogen bonds) hydroxyl groups of small propyl alcohol clusters are also observed in this frequency range. Such free hydroxyl groups are inherent in molecules that act as proton acceptors in the formation of larger dimers or open clusters. According to [14], IR absorption bands of propanol dimers appear at frequencies of 3654, 3671, and 3675 cm$^{-1}$. In the spectrum in Fig. 4 all three of these bands are present (denoted by the letter D), although they have a relatively low intensity. This fact indicates the presence of a small (compared to monomers) amount of dimers in the sample under study. Note that dimers were also present in the gas phase of propanol, as evidenced by the analysis of the spectra of gaseous propanol in the same spectral range (see Fig.1).

Registered maxima of IR bands of propanol monomers and dimers in an Ar matrix in the spectral region of the stretching vibrations of free hydroxyl groups as well as their assignments are listed in Table 1.

Table 1 – Spectral bands of stretching OH vibrations of propanol monomers and dimers

| v, cv$^{-1}$ | Assignment |
|---|---|
| 3654 | dimer, an acceptor molecule |
| 3657 | Gg conformer |
| 3659 | Tg conformer |
| 3660 | Gg' conformer |
| 3663 | Tt conformer |
| 3665 | Gt conformer |
| 3670 | dimer, an acceptor molecule |
| 3675 | dimer, an acceptor molecule |

Moreover, in the spectrum in Fig. 4 one can notice other low-intensity IR absorption bands, for example, at frequencies of 3643, 3644, 3648 and 3650 cm$^{-1}$, which most likely belong to clusters consisting of three or more molecules. Note that a band at 3647 cm$^{-1}$ was also registered in gaseous propanol and was attributed to the vibrations of trimers. Thus, it can be concluded that both in gaseous propanol and during its isolation in a low-temperature argon matrix, not only individual propanol molecules in different conformations exist there, but also hydrogen-bonded clusters consisting of two, three and more molecules are formed.

It is interesting to consider how the conformational composition of the sample under study changes with an increase in the temperature of the argon matrix. In fig. 5 the IR absorption spectrum of propanol in an argon matrix at 35 K in the spectral region of stretching vibrations of the free hydroxyl group is presented.

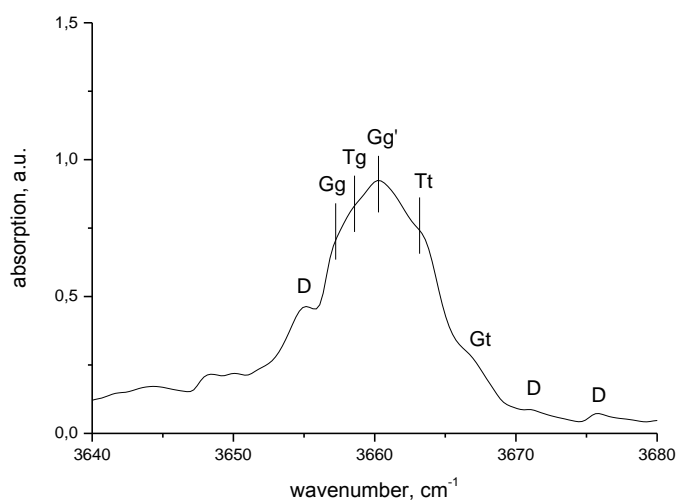

Fig. 5 – IR absorption spectrum of propanol isolated in an argon matrix at 35 K in the region of OH stretching vibrations

As can be seen from Fig. 5, with an increase in the temperature of the argon matrix to 35 K, the contributions of all propanol conformers remain in the absorption band of the free OH group, but the intensity of the Tt conformer absorption band noticeably increases compared to the other bands. This fact indicates a redistribution of the number of different conformers in the sample during its heating: the percentage of conformers with gauche orientation of the hydroxyl group decreases and the proportion of Tt conformers increases. This redistribution is most likely due to relaxation transitions from the Tg and Tg 'conformations to the Tt conformation, since the calculated height of the energy barrier between the Tt, Tg, and Tg' conformers is about 7 kJ/mol.

*Spectral region of stretching C – H vibrations*

Fig. 6 presents IR absorption spectrum of propanol in an Ar matrix at 20 K in the spectral region of the stretching C – H vibrations. In this region, symmetric and antisymmetric stretching vibrations in $CH_3$ group at 2886 and 2973 cm$^{-1}$ are the most intense, similarly to the picture observed for the gas phase (see Fig. 3). Between these two vibrations, there are symmetric stretching vibrations of $C_\beta H_2$ group. In contrast to the corresponding absorption bands in the gas phase, for which a strong overlap is observed (Fig. 3), the matrix isolation method allows the bands of different conformers to be separated.

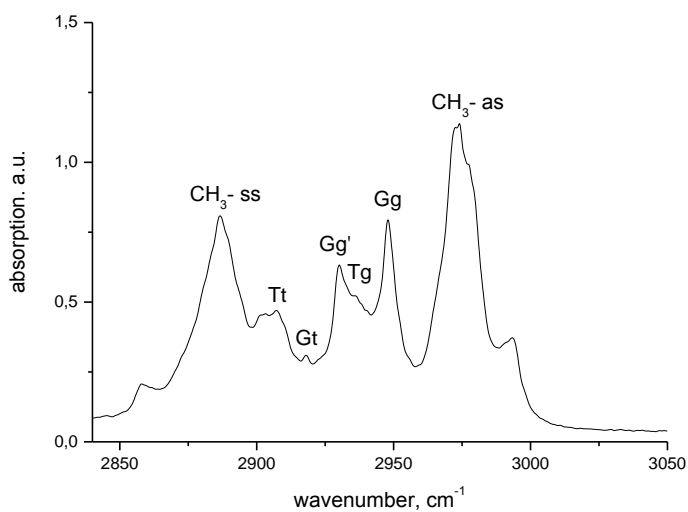

Fig. 6 – IR spectrum of propanol trapped in an Ar matrix at 20 K in the spectral region of stretching C – H vibrations

In Fig. 6 one can distinguish five bands of $C_\beta H_2$ vibrations corresponding to five different propanol conformers [16, 18]: the band of Tt conformer at 2907 cm$^{-1}$, the band of Gt conformer at 2918 cm$^{-1}$, the band of Gg' conformer at 2930 cm$^{-1}$, the one of Tg conformer at 2936 cm$^{-1}$ and of Gg conformer at 2948 cm$^{-1}$. From the distribution of the relative intensities of these five bands, it can be concluded that conformers with gauche orientation of the hydroxyl group, i.e. conformers of the Xg/g' type, predominate in the sample under study. This conclusion completely coincides with the results of the analysis of the spectral region of OH vibrations for this sample, which additionally confirms the correctness of the conclusions made above.

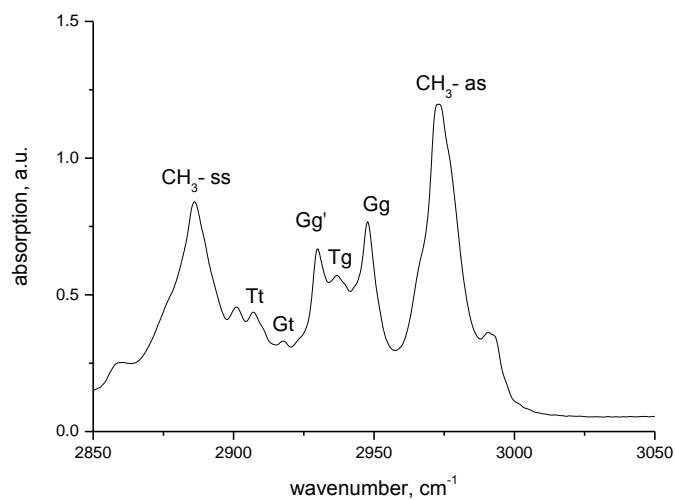

Fig. 7 – IR spectrum of propanol trapped in an Ar matrix at 35 K in the spectral region of stretching C – H vibrations

As the temperature of the matrix rises to 35 K, no noticeable changes in the spectrum in the spectral region of stretching CH vibrations occur. This is illustrated in Fig. 7, which shows the corresponding part of the IR absorption spectrum of propanol isolated in an argon matrix at a temperature of 35 K. The maxima of all the observed absorption bands remain at the same positions as at a temperature of 20 K. Only a slight narrowing of the bands of individual conformers can be noted, which leads to the formation of sharper maxima.

*Spectral region of skeletal vibrations*

Fig. 8 shows the IR absorption spectrum of propanol isolated in an argon matrix at a temperature of 20 K in the range of wavenumber from 800 to 1150 cm$^{-1}$, where the skeletal vibrations of the alkyl chain, i.e. stretching vibrations of C - C and C - O bonds, are located. In this part of the spectrum, rather narrow bands are observed, making it possible to assign them with high accuracy to various conformers of propanol.

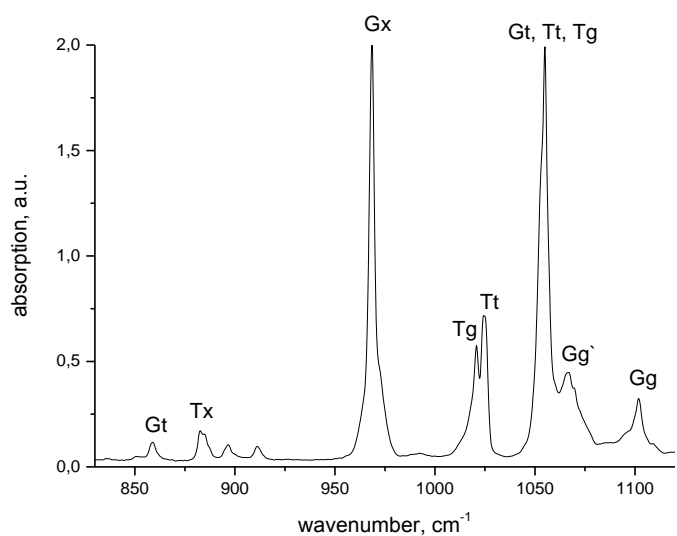

Fig. 8 – IR absorption spectrum of propanol trapped in an Ar matrix at 20 K in the spectral region 830 – 1120 cm$^{-1}$

At 859 cm$^{-1}$ a band of Gt conformer is observed, and at 883 cm$^{-1}$ - of a mixture of Tx conformers, similar to that in gaseous propanol (see Fig. 2).

One of the most intense absorption bands in this part of the spectrum at a frequency of 968 cm$^{-1}$ corresponds to the vibrations of the conformers of Gx group. This assignment was made based on a comparison of the results of quantum-chemical simulation of the IR absorption spectra of various propanol conformers performed by the B3LYP/6-311+G(d), MP2/6-311+G(d) and MP2/aug-cc-pVTZ methods with the experimentally recorded spectrum of propanol in an argon matrix [14]. Calculations showed the presence of this band in the spectra of the conformers Gt, Gg, and Gg ', and its absence in the spectra of the conformers Tt and Tg. Another band of approximately the same intensity was recorded at a frequency of 1055 cm$^{-1}$. According to the same calculations, vibrations at this frequency are present in the spectra of the Tt, Tg, and Gt conformers.

Using the results of quantum chemical calculations and experimental studies from [14], we assign the bands at 1020 and 1024 cm$^{-1}$ to the vibrations of Tg and Tt conformers, respectively. At a frequency of 1066 cm$^{-1}$, the absorption band of Gg' conformer is observed, and at a frequency of 1102 cm$^{-1}$ - of Gg conformer.

After analyzing the ratio of the intensities of the IR absorption bands of various conformers in this part of the spectrum, it can be concluded that Gx-type conformers prevail in number over Tx-type conformers. In addition, there are slightly more Tt conformers than Tg conformers, and the number of Gg structures is slightly less than Gg '.

When the sample was heated to 35 K, no significant changes in the spectral pattern were observed. Fig. 9 shows the same region of the IR absorption spectrum of propanol in an argon matrix, but at a temperature of 35 K. It can be seen that the only difference from the spectrum at 20 K is the practically equal intensities of the bands corresponding to the Tt and Tg conformers. In addition, an insignificant increase in absorption on the right wing of the Gx band and a splitting of the Gg' band can be observed, which may be associated with the appearance of additional bands in the spectrum as a result of the formation of larger propanol clusters with an increase of the sample temperature.

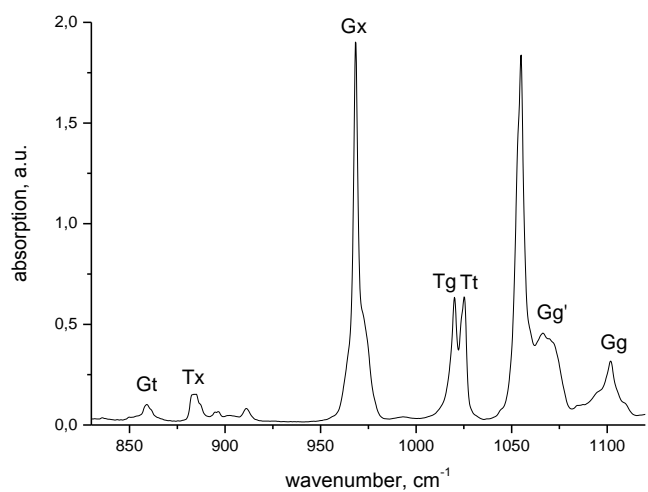

Fig. 9 – IR absorption spectrum of propanol trapped in an Ar matrix at 35 K in the spectral region 830 – 1120 cm$^{-1}$

*Spectral region of bending OH vibrations*

Let us consider separately the spectral region 1100 - 1300 cm$^{-1}$, where bending OH vibrations of propanol are manifested. The corresponding part of the recorded IR absorption spectra of propanol isolated in an argon matrix at temperatures of 20 K and 35 K is shown in Fig. 10 and 11, respectively. As can be seen, vibrations of all five propanol conformers are recorded in this region, the assignment of which was carried out on the basis of our quantum-chemical calculations, as well as the results of modeling presented in [14].

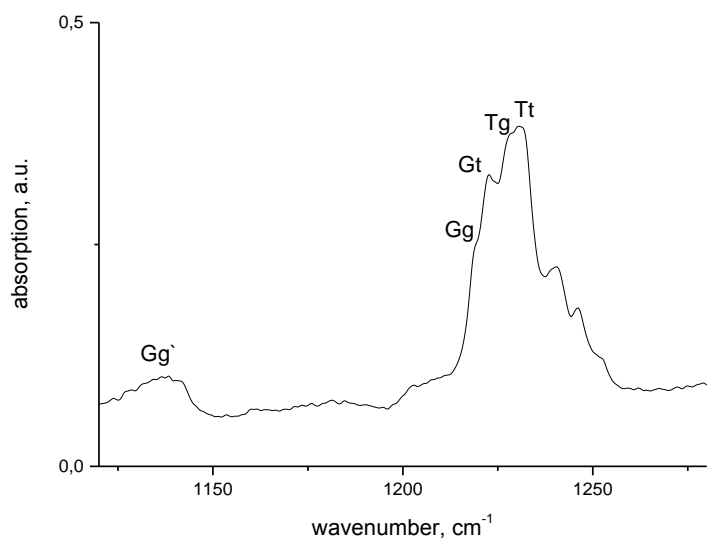

Fig. 10 – IR absorption spectrum of propanol in an Ar matrix at 20 K in the spectral region of bending OH vibrations

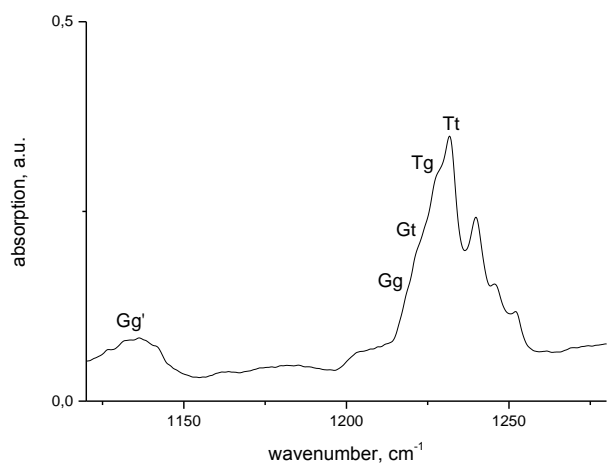

Fig. 11 – IR absorption spectrum of propanol in an Ar matrix at 35 K in the spectral region of bending OH vibrations

Quantum-chemical simulation of IR absorption spectra of various propanol conformers shows that bending OH vibrations of Tt conformer appear at a frequency of 1232 cm$^{-1}$, and ones of Tg conformer – at a frequency of 1228 cm$^{-1}$. The corresponding absorption band of the Gt conformer is located at a frequency of 1222 cm$^{-1}$, and the vibrations of the Gg conformer manifest as a shoulder of this band at 2019 cm$^{-1}$. For Gg' conformer the calculated frequencies of bending OH vibrations are somewhat lower - about 1130 cm$^{-1}$.

All the bands listed above are observed at both temperatures, but with slightly different intensity distributions. In addition, the spectrum contains bands at frequencies of 1240, 1246, and 1252 cm$^{-1}$. Since the intensity of these bands increases with increasing temperature, it can be assumed that they belong to propanol clusters.

**Conclusions**

Conformational analysis of the experimentally recorded IR absorption spectra of propanol in gaseous state and in a low-temperature argon matrix, carried out for different spectral ranges, showed that the conformational composition of the samples in these two cases is different.

Analyzing the region of OH stretching vibrations in gaseous propanol, the predominance of Gt structures, as well as the presence of the Gg and Gg' structures, was observed. This composition of the sample in the gas phase is also confirmed by the results of the analysis of the spectral region of bending OH vibrations, which indicate the predominance of the Gt form over Tt, and the region of skeletal vibrations, according to which conformers of the Gx type prevail over those of Tx.

When propanol is isolated in an argon matrix, a redistribution occurs between the ratios of conformers in the sample. Analysis of the range of OH stretching vibrations shows the predominance of the Tg conformer, but the number of Gg and Gg' conformers is not much less. The slight predominance of structures of the Tx type over Gx is also evidenced by the distribution of the intensities of the recorded absorption bands in the spectral intervals of bending OH vibrations and skeletal vibrations of the alkyl chain. The spectral region of stretching CH vibrations turned out to be the least informative both in the case of a gaseous sample and in the case of matrix isolation.

Thus, the results of our studies show that gaseous propanol contains the largest number of Gt conformers, which are the most stable of the five possible propanol conformers according to a number of quantum mechanical calculations [8, 14, 21]. The propanol molecules isolated in a low-temperature argon matrix are influenced by the environment; therefore, in this case the most energetically favorable form is Tg conformer, which prevails in percentage.